# High-Transmission Mid-Infrared Bandpass Filters Using Hybrid Metal-Dielectric Metasurfaces for $CO_2$ Sensing

Amr Soliman*, C Williams, Richard Hopper, Florin Udrea, Haider Butt, Timothy D. Wilkinson

**Amr Soliman, Richard Hopper, Florin Udrea, Timothy D. Wilkinson** - Electrical Engineering Division, Department of Engineering, University of Cambridge, 9 JJ Thomson Avenue,
Cambridge CB3 0FA, UK, https://orcid.org/0009-0000-6040-0039

**Calum Williams** - Department of Physics, University of Exeter, Stocker Rd, Exeter, EX4 4QL United Kingdom

**Haider Butt** - Department of Mechanical and Nuclear Engineering, Khalifa University of Science and Technology, Abu Dhabi, United Arab Emirates

**ABSTRACT:** Mid-infrared (MIR) spectroscopy is a powerful technique employed for a variety of applications, including gas sensing, industrial inspection, astronomy, surveillance, and imaging. Thin-film narrowband interference filters—targeted to specific absorption bands of target molecules—are commonly deployed for cost-effective MIR sensing systems. These devices require complex and time-consuming fabrication processes. Also, their customization on the micro-scale for emerging miniaturized applications is challenging. Plasmonic nanostructure arrays operating in reflection and transmission modes have been developed for MIR. However, they experience undesirable characteristics, such as broad spectra and low reflection/transmission efficiencies. All-dielectric metasurfaces have low intrinsic losses and have emerged as a substitute for plasmonic metasurfaces in MIR spectroscopy. Nevertheless, they typically operate only in reflection mode. In this work, we present a hybrid metal-dielectric metasurface for MIR spectroscopy operating in transmission mode. The metasurface is composed of germanium (Ge) atop aluminum (Al) cylinders, and we show that the transmission response arises because of the hybridization of modes arising from the Ge and the Al structures. The presented metasurface has a high transmission efficiency of 80 % at $\lambda = 2.6$ µm, and a narrow full-width-at-half-maximum of 0.4 µm. We show numerical simulations, successful fabrication using a straightforward fabrication method, and deployment as the in-line optical filter in a $CO_2$ gas detection with a limit of detection of ~0.04% (a few hundred ppm). Our work demonstrates the potential for hybrid metasurfaces as in-line gas sensing optical filters in MIR spectroscopy.

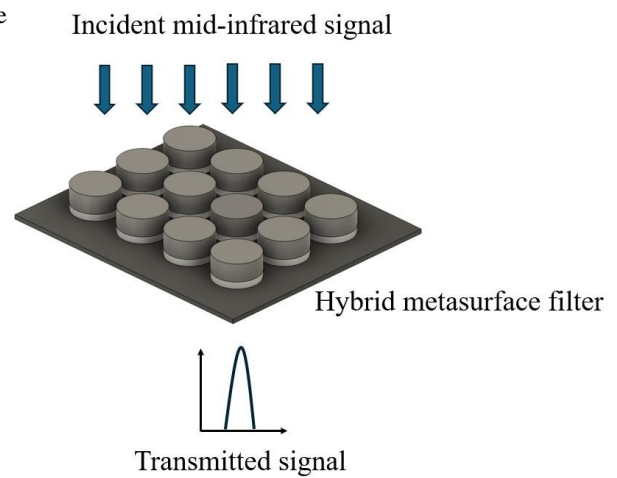

Mid-infrared (MIR) spectroscopy plays an essential role in sensing applications due to its ability to detect characteristic molecular absorption bands resulting from the intrinsic vibrational modes of chemical bonds. This technique enables direct analysis of molecular structures with absorption features distinct to the MIR spectral range [1]. MIR spectroscopy has numerous applications across astronomy, surveillance, and industrial inspection [2-5].

Many MIR spectroscopy devices are based on Fourier-transform-IR (FTIR) approaches [6, 7] and linear variable optical filters (LVOFs) [8, 9]. These approaches are commonly employed in applications where the chemical species are known in advance, such as gas detection and multiplexing [8-12]. However, these approaches can be costly and challenging to customize or miniaturize. Alternative methodologies such as single wavelength or multispectral filters tuned to the desired chemical absorption peaks may be utilized [13, 14]. For example, Fabry-Perot (FP) based bandpass filters are typically used [15-17]. FP-bandpass filters consist of two distributed Bragg reflectors (DBRs) separated by a well-defined central resonant cavity. The optical thickness of the central cavity specifies the center wavelength (CWL) while the DBRs determine the reflection stop-band [17]. Despite providing high transmission efficiency and narrow passbands, FP filters require lengthy microfabrication (physical vapor deposition) processes to deposit different layers, and spatially-variant customization (or miniaturization to the chip scale) can be challenging. Moreover, MIR FP filters exhibit a narrow stop band due to the index contrast of the available MIR transparent materials resulting in crosstalk between the different spectral responses which affects their filtering effect [16].

Over recent years, plasmonic/all-dielectric metasurfaces have been used in MIR spectroscopy, from in-line optical filters to surface-enhanced scattering devices [18-20]. Metasurfaces are arrays of nanostructures with a sub-wavelength thickness that can be used to modulate both the amplitude and the phase of the incident light to achieve sophisticated optical functions in compact form factors such as negative refraction and 'perfect' lenses [21]. Metasurfaces can be fabricated from metals (plasmonic metasurfaces) or dielectrics (all-dielectric metasurfaces). Plasmonic and all-dielectric metasurfaces can provide a spectral filtering effect through the excitation of electric and magnetic resonances. Plasmonic metasurfaces operating in both reflection and transmission modes have been employed in MIR spectroscopy. When the plasmonic resonance aligns spectrally with the characteristic molecular absorption, the interaction between the molecule and the resonator can modify the frequency or intensity of the resonance, enabling the identification of molecular. This concept, known as surface-enhanced infrared absorption (SEIRA), has been implemented using various plasmonic-based devices [22-24]. However, plasmonic metasurfaces experience different problems, such as broad spectra and low reflection/transmission efficiencies, because of the high ohmic losses associated with the plasmonic metals [25]. Also, the spectra of plasmonic nanostructures have multiple resonant features due to the excitation of different plasmonic modes. This can be undesirable in some applications, such as sensing that requires a spectrum with only one or two sharp features to track the change in the spectral position of the spectral features [26].

All-dielectric metasurfaces have emerged as alternatives to plasmonic metasurfaces due to their low intrinsic loss [25, 27]. All-dielectric metasurfaces have been used in MIR spectroscopy [19, 20], however, all-dielectric metasurfaces typically operate in reflection mode only because they can only generate transmission valleys (notches) [28]. Additionally, all-dielectric-based designs exhibit lower field enhancement affecting their performance in gas sensing applications which require high signal throughput from low concentration levels [29]. Hence, there is a need for a miniaturizable MIR spectroscopy filtering technique that operates in transmission mode with high transmission efficiency, narrow full-width-half-maximum (FWHM), and localized field enhancement.

Hybrid metal-dielectric metasurfaces are attracting interest due to their ability to mitigate the high ohmic absorption losses associated with plasmonic metals, while also avoiding the low field enhancement typically observed in all-dielectric metasurfaces. In effect, hybrid dielectric-metal metasurfaces combine the advantages of both plasmonic metals and dielectric metasurfaces [29]. This results in high transmission efficiencies and controllable transmission responses. Hybrid metasurfaces also have an ultra-narrow bandwidth (high Q-factor [30]) with minimal cross-talk. Hybrid metasurfaces have been utilized to produce high-performance structural colors in transmission mode [28]. The interaction between Wood's anomaly in Al nanostructures and the Mie lattice resonance in $Si_3N_4$ nanostructures generated magnetic dipole and electric quadrupole resonances, achieving a high transmission resonance with an efficiency of approximately 70%. Despite the use of hybrid metasurfaces in various applications, including higher-order harmonic generation [31], high directivity nanoantennas [32], and fluorescence enhancement [33, 34], there is a scarcity of studies on metasurfaces operating in transmission mode across the MIR wavelength range.

In this work, we propose a hybrid metal-dielectric (Al-Ge) metasurface for MIR spectral bandpass filtering and deploy it as an in-line optical filter for $CO_2$ gas sensing. We design the filter to exhibit a transmission mode at a $CO_2$ vibrational absorption mode and show a transmission efficiency of 80% at $\lambda = 2.6$ μm, with FWHM of 0.4 μm. Through computational modeling, we demonstrate that the hybridization between the electric dipole (ED) of the Al metasurface and the ED, magnetic dipole (MD), electric quadrupole (EQ), and magnetic quadrupole (MQ) of the Ge metasurface results in high transmission at $\lambda = 2.6$ μm, attributed to the excitation of ED, MD, EQ, and MQ modes within the hybrid metasurface. We further demonstrate that the spectral response of the reported filter can be adjusted across the MIR wavelength range by modifying its geometrical parameters, facilitating the development of bandpass filters with tunable center wavelengths (CWL). The optical filter is numerically simulated, nanofabricated, characterized, and used to achieve a $CO_2$ limit of detection (LOD) down to ~0.04%.

Despite their superior performance, the fabrication process of the hybrid metal-dielectric metasurface is challenging due to their structural complexity [29, 35]. To address these issues, a straightforward fabrication process was developed based on one lithography step followed by the deposition of metals/dielectrics and, finally, the typical lift-off process to define the metasurface structures. This process can be applied to fabricating complex structures.

**RESULTS AND DISCUSSIONS**

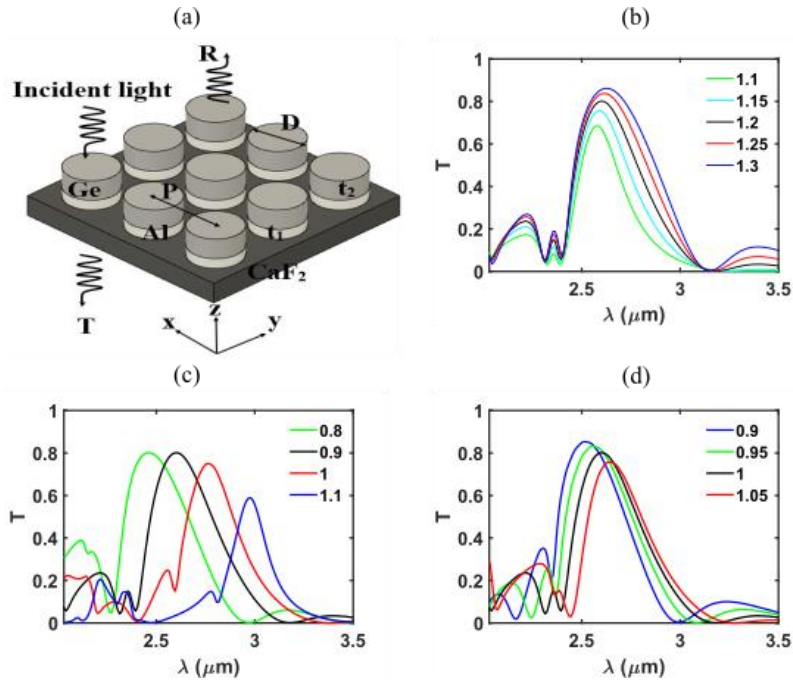

Figure 1. (a) A schematic representation of the hybrid metal-dielectric metasurface showing different geometrical parameters. (b) (c) and (d) Variation of $T$ as a function of $\lambda$ for different $P$, $D$, $t_1$, and $t_2$, respectively, showing the effect of changing different geometrical parameters on the spectral response of the metasurface.

**Materials and Design.** Figure 1(a) shows a schematic of the hybrid metal-dielectric metasurface with the respective geometrical parameters. The geometrical parameters are as follows: $P$ is the period (center to center distance), $D$ is the diameter of the pillars, $t_1$ and $t_2$ are the thicknesses of the Au and Ge layers, respectively. Each unit cell is composed of Ge atop Al cylinders on a calcium fluoride ($CaF_2$) substrate. $CaF_2$ was chosen as the substrate material due to its comparatively low refractive index (1.4) and negligible losses in the MIR [36]. Al is chosen as the plasmonic metal as it is highly abundant in nature, compatible with CMOS processes, and cost-effective [29], while Ge is selected as the dielectric because it has a high refractive index and low losses (extinction) in the MIR region [37]. The presented design was chosen because of its straightforward experimental realization, minimal polarization sensitivity, high incident angle tolerance, and the well-known characteristics of its building blocks [38-40].

**Design Concept and Parametric Sweeps.** The incident light on hybrid metal-dielectric metasurfaces excites two modes: the plasmonic mode in the metallic part and the Mie resonances in the dielectric part. The final response is the coupling between these two modes [28]. With the proper manipulation of the geometrical parameters of the hybrid metasurface, we can get the desired transmission response in terms of its spectral position, the transmission efficiency, and the FWHM of the transmission peak. Hence, parametric sweeps for all the geometrical parameters shown in Figure 1(a) were performed to achieve the highest transmission efficiency, the narrowest FWHM, and minimum secondary spectral features at a CWL of $\lambda = 2.6$ µm. Throughout the parametric sweeps, one parameter was varied while all other parameters were held constant. In all simulations, the input plane wave was an x-polarized E-field. Periodic boundary conditions were used in the x and y directions in order, and perfectly matched layers (PMLs) were used in the z-direction. The mesh size was kept at 5 nm for all simulations. The complex dispersive refractive indices for Al were calculated from the multi-coefficient model, and the refractive indices of thin film Ge were measured using an in-house ellipsometer (Model: Woollam Ellipsometer M-2000). The ellipsometer measurements were performed at incidence angles of 45°, 50°, 55°, and 60°. The measured spectra were fitted to theoretical spectra generated by a multilayer model. The refractive indices were determined using the Genosc model [41]. After fitting, the calculated spectra were well-matched with the measured ones at all wavelengths and angles. Then, the optical constants of the deposited Ge were obtained.

Figure 1(b) shows the variation of the transmission efficiency as a function of $\lambda$ for different periods ($P$). The figure depicts that increasing $P$ leads to higher transmission efficiency, an increased FWHM of the transmission peak, and redshifts the central wavelength of the peak. As shown in Figure 5(a), the maximum transmission efficiency increased from 68% at $\lambda = 2.57$ µm to 86% at $\lambda = 2.63$ µm by increasing $P$ from 1.1 µm to 1.3 µm, whereas the FWHM increased from 0.36 µm at $P = 1.1$ µm to 0.45 at $P = 1.3$ µm. Arranging hybrid metasurfaces into periodic arrays induces proximity resonances between adjacent unit cells. Consequently, adjusting $P$ alters the coupling effect between the unit cell resonances, thus impacting both the transmission efficiency and FWHM [42]. By increasing $P$, the gap between the pillars increases, hence the coupling between the neighboring pillars decreases, resulting in gradually widening the FWHM of the spectra [28, 42]. Moreover, the reduced coupling between the neighboring unit cells can also increase the transmission at undesired wavelengths (around $\lambda = 2.2$ µm and 3.3 µm). Ultimately, a period of $P = 1.2$ µm was selected for the final design since it achieves high transmission equals 80% at $\lambda$ at the desired $\lambda$ (2.6 µm) associated with a strong $CO_2$ absorption mode, a narrow FWHM (0.4 µm), and minimum transmission at $\lambda = 2.2$ µm and 3.3 µm. Recently, precise manipulation of $P$ has also been utilized to significantly enhance the coupling between the unit cells, resulting in improving the transmission efficiency of the structural colors [28].

The effect of varying the diameter ($D$) on the transmission was also investigated, as shown in Figure 1(c). The figure shows that increasing $D$ from 0.8 µm to 1.1 µm redshifts the transmission peak. This happens because changing the diameter of the pillars while maintaining a constant period modifies the filling factor of the hybrid metasurfaces. According to the principles of constructive interference and equivalent refractive index, adjusting either the thickness or the filling factor of the hybrid metasurfaces results in varying effective optical path differences, resulting in different transmission peaks [43, 44]. This concept has been used before to generate distinct transmitted colors in the visible regime under white-light illumination [44]. The chosen $D$ for the final design was 0.9 as it has the highest transmission efficiency (80%) at $\lambda = 2.6$ µm and FWHM equal 0.4 µm.

The thickness of Al ($t_1$) has a minor impact on the transmission efficiency without affecting the central wavelength. However, it slightly affects the transmission efficiency at $\lambda = 2.2$ µm and 3.3 µm. The thickness $t_1 = 0.085$ µm was selected for the final design because it has the highest transmission efficiency (80%) at $\lambda = 2.6$ µm. The variation in transmission efficiency as a function of $\lambda$ for different Ge thicknesses ($t_2$) is presented in Figure 1(d). The figure demonstrates that increasing $t_2$ from 0.9 µm to 1.05 µm results in a redshift of the transmission efficiency. This redshift is attributed to the increased effective optical path that occurs with the increased thickness [44]. Furthermore, increasing $t_2$ from 0.9 µm to 1.05 µm decreases the transmission efficiency from 85% at $\lambda = 2.52$ µm to 75% at $\lambda = 2.65$ µm, while the FWHM decreases from 0.44 µm to 0.38 µm. The highest transmission efficiency (80%) at the target wavelength of $\lambda = 2.6$ µm was achieved at $t_2 = 1$ µm, thus it was selected for the final design.

**Advantages and Physical Mechanism of Hybrid Metasurfaces.** To highlight the merits of the hybrid metal-dielectric metasurfaces and the advantages they added to the response, the transmission responses of the Al only, Ge only, and hybrid Al-Ge metasurfaces were calculated at the optimized geometrical parameters. Furthermore, the multipolar expansions and the scattering cross section (SCS) were also calculated to interpret the physical mechanism of their responses, as the transmission spectra are governed by scattering [45].

We simulated the multipole expansion and the SCS using an open-source MATLAB code (MENP, multipole expansion for nanophotonics) [46]. MENP is a MATLAB program used for the computation of the electric dipole (ED), magnetic dipole (MD), electric quadrupole (EQ), and magnetic quadrupole (MQ) followed by the calculation of the SCS. It calculates the electric fields excited in the metasurface using Lumerical FDTD. Then, the calculated fields are packaged into .mat file format together with refractive index (n(x, y, z, f)) and one-dimensional arrays of axes (x, y, z, f). Next, the fields are passed to MENP which firstly converts the electric field distributions into current density. Finally, MENP uses the calculated current density to compute the four modes (ED, MD, EQ, and MQ) followed by the SCS.

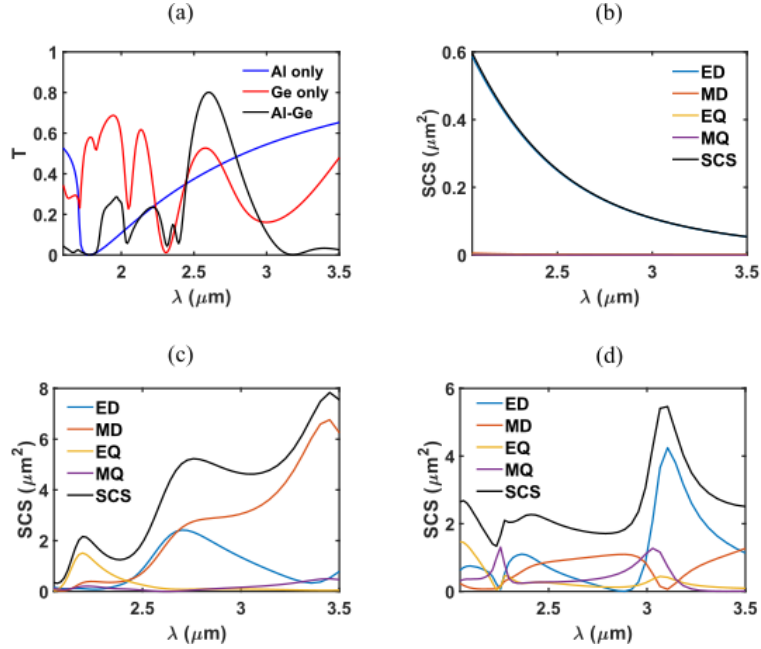

Figure 2. (a) Variation of *T* as a function of λ for the Al only, the Ge only, and the hybrid metasurfaces. The figure shows the effect of using the hybrid metasurface on the spectral response. The multipole expansions of the SCS in terms of ED, MD, EQ, and MQ for (b) the Al-only, (c) the Ge-only, and (d) the hybrid metasurface.

Figure 2(a) presents the transmission of the Al only, the Ge only, and the hybrid metasurfaces. The figure demonstrates that adding Ge to Al has increased the transmission at the target wavelength of 2.6 μm while suppressing the transmission elsewhere. The figure also shows that the transmission response of the Ge only metasurface has multiple peaks whereas the transmission response of the Al only metasurface does not have any sharp peaks. This is undesirable in sensing applications that require only one or two sharp features across the operating wavelength range [26].

To comprehend the physical origin of the resonant mode and determine the contributions of Al and Ge metasurfaces to the resonant response of the hybrid metasurface, the multipole expansions and the SCS of the single and hybrid metasurfaces were calculated. Figures 2(b), (c), and (d) demonstrate the SCS of the Al only, the Ge only, and the hybrid metasurfaces, respectively. Figure 2(b) indicates that the Al-only metasurface features a dominant, broad electric dipole (ED) resonance. Due to the non-directional nature of electric dipolar scattering [47], the Al-only metasurface demonstrates low transmission with no sharp peaks, as depicted in Figure 2(a). Figure 2(b) shows that the Ge-only metasurface exhibits additional higher-order modes such as MD, EQ, and a minor contribution from the MQ, consistent with previous studies indicating that the dielectric component excites higher-order modes [26, 28]. When combining Al with Ge, the hybridization between the ED of the Al only metasurface with the ED, MD, EQ, and MQ of the Ge only metasurface resulted in a high transmission at λ = 2.6 μm due to the excitation of ED, MD, EQ, and MQ modes within the hybrid metasurface. As shown in Figure 2(d), the response of the hybrid metasurface is shifted to around λ = 3.1 μm instead of λ = 2.6 μm. The shift can be attributed to the approximations done while calculating the SCS and the multipole expansion [46].

Figure 3(a) shows the E-field distribution of the hybrid metasurface in the *yz* plane (associated with Figure 1(a)). As shown in the Figure, the E-field is concentrated around the unit cells of the metasurfaces showing a high sensitivity to the surrounding environment.

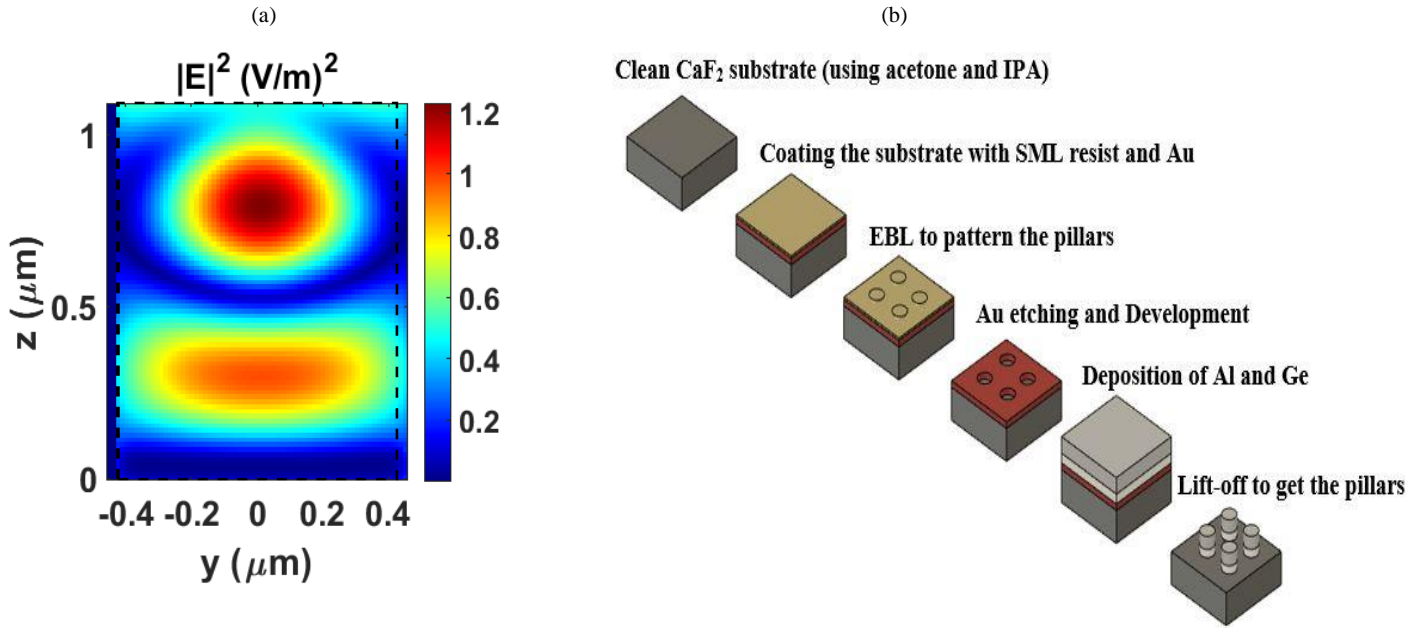

Figure 3. (a) The E-field for the hybrid metasurface at λ = 2.6 μm in the yz plane. (b) A schematic shows different steps for the fabrication process of the hybrid metasurface. The hybrid metasurface is fabricated using EBL, deposition of Al and Ge followed by lift-off.

**Experimental Platform.** Although they exhibit superior performance, hybrid metal-dielectric metasurfaces present significant fabrication challenges owing to their complex structural configuration [27]. Previously reported hybrid metasurfaces were fabricated using either one [27, 28] or two lithography steps [35]. Two-step lithography processes increase both fabrication time and cost. A precise alignment between the two lithography procedures is also needed, which can be challenging, especially for high-resolution patterning. Alternate designs fabricated using a single lithography step often necessitate reactive ion etching (RIE). The thickness of metasurfaces significantly impacts the effective optical path of the excited mode [44], thus playing a crucial role in determining the spectral position of the excited mode. Metasurfaces operating in the MIR region require increased thicknesses of the resonators, due to the required long optical path. Achieving anisotropic etching for such a high thickness is challenging and typically requires deep reactive ion etching (DRIE) [48]. DRIE utilizes a cyclic process alternating between an etching step using reactive gases like $SF_6$ and a passivation step using gases like $C_4F_8$. During the etching step, the reactive gas removes the etched material, while the passivation step deposits a protective polymer film on the sidewalls of the etched features. This cyclic process results in highly anisotropic and deep etching profiles for high aspect ratio structures [49, 50]. However, there are challenges associated with DRIE. These challenges include etching lag [51, 52]. RIE lag reduces the etch rate for the pillars with high aspect ratios because the etchant gas dynamics during the etching and the passivation steps limit and obstruct the gas from reaching the bottom of the pillar. This results in different etching rates between the top and the bottom of the pillar affecting the final etched structure and consequently significantly influencing its performance. Another challenge is the low selectivity (ratio between the etch rates of the mask and Ge) of DRIE [52]. To overcome these challenges, a straightforward fabrication process was employed, involving a single lithography step, followed by the deposition of metals and dielectrics, and finally, the standard lift-off process to define the metasurface structures.

Figure 3(b) illustrates the metasurface fabrication process. Initially, a $CaF_2$ substrate is cleaned using acetone and isopropyl alcohol (IPA). The substrate is then coated with SML electron beam resist (EM Resist Ltd.) for electron beam lithography (EBL) patterning. SML resist can simultaneously pattern high-resolution and high-aspect ratio patterns. SML can be used at low acceleration voltages without needing proximity effect correction. It has three times more sensitivity than the commonly used electron beam resist PMMA. SML has been designed to be compatible with the standard PMMA process [53]. SML enabled us to fabricate the hybrid metal-dielectric metasurfaces with a much simpler approach. Instead of using a two-step EBL procedure, one lithography step was used to pattern the hybrid metasurfaces, followed by Al and Ge deposition. Then, the lift-off process was used to define the final structures. The reported fabrication process reduces the time and cost of fabricating the hybrid metasurfaces by eliminating one lithography step. Moreover, lift-off replaced etching with its associated problems. Compared to SML, PMMA has high density, which causes a relatively large amount of scattering for the e-beam through the resist, blurring it by the time it reaches the bottom. This will significantly increase the resist undercut where the resist becomes detached from the surface of the substrate, especially in the narrow gaps between the features, which is sufficient to collapse the pattern and clear all the resist. This limits the aspect ratio of large features in thick resists. That is why low-density/low-scattering resists such as SML were designed to minimize scattering and maintain the undercut, which allows for high aspect ratio patterns. Additionally, viscous resists like PMMA can be difficult to coat uniformly, especially for thick resists with slow spinning speeds. SML solves this issue by having a lower density than PMMA [54-56].

The soft bake is a crucial step in resist preparation following the spin coating of the resist onto the substrate. Its primary function is to remove residual solvent from the resist, thereby enhancing adhesion to the substrate and eliminating bubbles within the resist film [57, 58]. Bubbles, which may arise from transport, dispensing, or refilling of the resist, can serve as initiation points for cracks in the resist film [59]. Bubbles, potentially arising from the transport, dispensing, or refilling of the resist, can serve as initiation points for cracks in the resist film.

Initially, the soft bake for the SML resist was performed at the manufacturer's recommended temperature of 180 °C for 2 minutes [53]. However, following the EBL exposure and the development, the resist film exhibited cracks (Figure 4a), indicating that the soft bake temperature was insufficient, thereby leading to crack formation. These cracks pose potential issues for the subsequent lift-off process, necessitating optimization of the soft bake temperature. Various soft bake temperatures were investigated. The chosen soft bake temperatures were selected based on their prior successful use with PMMA resist, given that the SML resist is designed to be compatible with the standard PMMA process. Initially, a soft bake at 150 °C for 5 minutes was tested, but cracks were still observed in the resist film post-EBL exposure and development as Figure 4(b) depicts. A two-step soft baking approach was then employed: the sample was initially baked at 160 °C, followed by an increase to 180 °C, with a gradual decrease in temperature to avoid thermal shocks, over a total period of 6 minutes, however, Figure 4(c) shows still cracks in the resist. Ultimately, soft baking at 160 °C for 4 minutes resulted in a crack-free resist film (Figure 4d).

The fabrication process was completed for devices baked at various soft bake temperatures to assess the impact of cracks on the final devices Tilted SEM images of the fabricated devices are shown in Figures 4(e), (f), (g), and (h). The cracks shown in Figures 4(a), (b), and (c) led to poor adhesion between the deposited Al and Ge films, complicating the lift-off process. During lift-off, acetone stripped off parts of the deposited Ge film, leaving only the Al pillars, as seen in Figures 4(e), 4(f), and 4(g). Conversely, the device baked at 160 °C for 4 minutes exhibited no cracks as shown in Figure 4h, indicating that this soft baking condition effectively eliminated bubbles in the resist film and prevented crack formation. Consequently, a soft bake temperature of 160 °C for 4 minutes was adopted for subsequent processes.

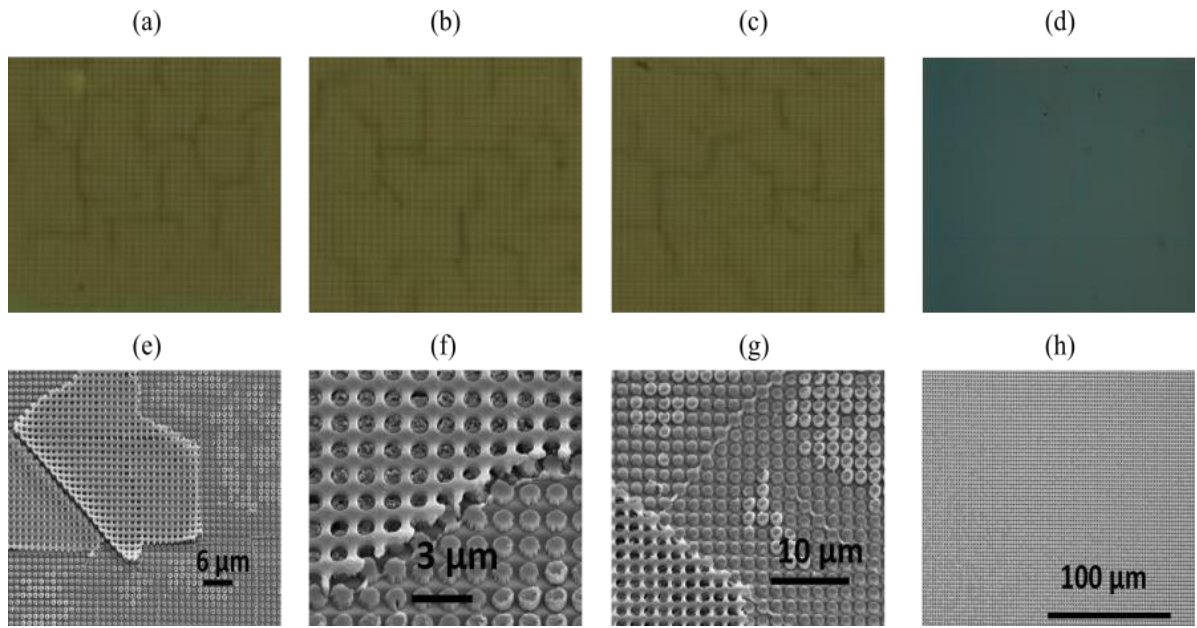

Figure 4. (a), (b), (c), and (d) Optical microscopy images of samples baked at different soft bake temperatures (a) 180 °C for 2 minutes, (b) 150 °C for 5 minutes (c) 160 °C then ramp to 180 °C (d) 160 °C for 4 minutes. The images were taken after EBL exposure and development. (e), (f), (g), and (h) SEM images of the samples fabricated using soft bake temperatures (a), (b), (c), and(d), respectively. The figures show that the sample baked at 160 °C for 4 minutes has no cracks.

15 nm of Au was deposited on the SML resist to dissipate the charges during the EBL exposure. The resist was patterned by EBL using four different doses 830, 870, 910, and 950 $\mu C/m^2$. After EBL exposure and before development, the 15 nm Au layer was removed using a wet-etchant (30 seconds in gold etchant, standard from Sigma-Aldrich at room temperature). Next, the samples were developed in MIBK: IPA (1:3) for 45 seconds. After development, Al and Ge were deposited using a thermal evaporator (MiniLab 60, Moorfield Nanotechnology Ltd). Then, the substrates were left in acetone overnight for the lift-off process. Figure 5(a) shows different SEM images for the fabricated hybrid metasurfaces using EBL exposure dose 830 $\mu C/m^2$. The SEM image of the fabricated hybrid metasurface demonstrates a high degree of fabrication uniformity.

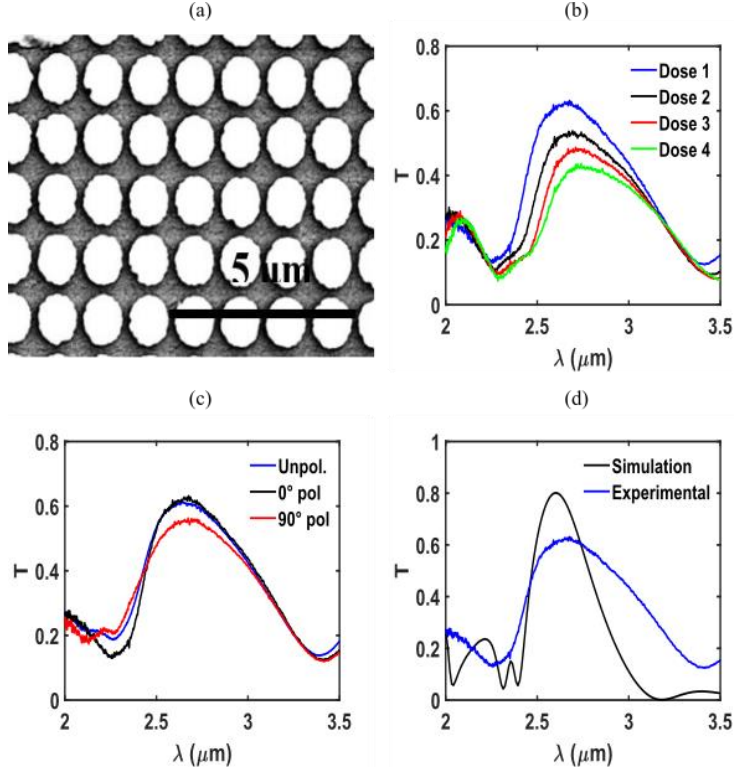

Figure 5. (a) SEM image of the fabricated hybrid metasurface using EBL exposure dose of 830 µC/m$^2$. The SEM image reveals a high level of uniformity in the fabrication process (b) Measured transmission spectra of metasurfaces patterned using different EBL exposure doses (Dose 1 = 830, Dose 2 = 870, Dose 3 = 910, and Dose 4 = 950 µC/m$^2$). (c) The measured transmission response of the hybrid metasurface fabricated using 830 µC/m$^2$ EBL exposure dose. The figure illustrates that varying the polarization has a minimal impact on the transmission response. (d) Comparison between simulated and measured results. The figure shows that the simulated and measured results have the same central wavelength.

**Optical Characterization.** The fabricated hybrid metal-dielectric metasurfaces were then optically characterized. The spectra of the fabricated metasurfaces were measured using an FTIR microscope (Shimadzu AIM-9000 FTIR Infrared Microscope). The measurements were performed in transmission mode. A bare CaF$_2$ substrate was first measured as a reference. Figure 5(b) presents the measured transmission responses of four hybrid metasurfaces under 0° light incidence. These metasurfaces were patterned using various EBL exposure doses, resulting in differing diameters. As shown in Figure 5(b), the central wavelength of all four filters is approximately 2.6 µm, with dose 1 and dose 2 exhibiting the highest transmission. Variations in diameters due to different EBL exposure doses led to differences in the transmission efficiencies of the four devices. Figure 5(c) displays the transmission response of the metasurface fabricated using EBL exposure dose 830 µC/m$^2$ under unpolarized, 0°, and 90° light incidence. The results indicate that changing the polarization has a minor effect on transmission efficiency without shifting the central wavelength, demonstrating that the hybrid metasurfaces are polarization-insensitive. The difference in transmission efficiency for different polarization states is attributed to shape deformation and minor shape variations across the *x* and *y* directions.

Figure 5(d) shows a comparison between the measured transmission efficiency of the device fabricated using dose 1 (830 µC/m$^2$) and the simulated transmission for the following parameters: $P = 1.2$ µm, $D = 0.9$ µm, $t_1 = 0.085$ µm and $t_2 = 1$ µm. It is worth noting that there is an agreement between the simulated result and the experimental measurement regarding the central wavelength, as they both have the same CWL. However, the measured result has a lower transmission efficiency (>60%) than the simulated one (>80%). This difference in transmission efficiency can be attributed to the surface roughness of the metasurface unit cells. Surface roughness produces extra scattering to the incident light and reduces the measured transmitted light [42]. Additionally, the FWHM of the measurement is greater than that of the simulated result. The resonance broadening occurs because of the fabrication variations over the measured area and surface roughness of the metasurface unit cells [29, 60, 61].

The surface roughness of the device fabricated using an EBL dose of 830 µC/m$^2$ was measured using an AFM. The root-mean-square roughness (RMS) roughness of one pillar measured over an area of 0.6 µm × 0.55 µm was equal to 46.4 nm. Similarly, another pillar, measured over an area of 0.5 µm × 0.5 µm, exhibited an RMS roughness of 46.6 nm, indicating significant surface roughness. The high surface roughness of each Al-Ge unit cell led to a scattering of the incident light, reducing the measured transmission efficiency [42]. Hence, the measured transmission is lower than the simulated one, as shown in Figure 5(d).

Furthermore, the high surface roughness of the unit cells results in an inhomogeneous broadening of the response [60, 61] as illustrated in Figure 5(d). Also, the fabrication defects over the sample area contributed to the inhomogeneous broadening of the measured result [28, 29]. In conclusion, the lower transmission efficiency and the broader response of the measurement can be attributed to the measured high surface roughness and fabrication defects.

The surface roughness can be significantly reduced by changing the deposition technique of Ge. For example, E-beam evaporation can produce excellent deposited Ge films compared to other methods such as thermal evaporator and sputtering [62]. This can improve the quality of the deposited films and subsequently the measurements.

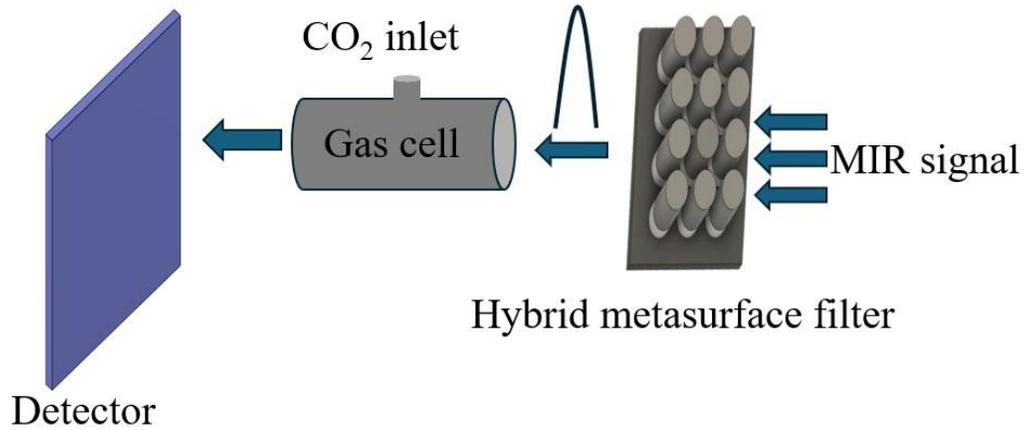

Figure 6. A schematic of the gas sensor system. The MIR signal is directed onto the hybrid metasurface filter. The transmitted signal is then injected to the gas cell, which is linked to a detector that measures the intensity of the signal before and after the introduction of $CO_2$.

The hybrid Al-Ge metasurface was tested in a proof-of-concept NDIR (nondispersive infra-red) gas sensor system to detect $CO_2$. A schematic of the gas sensor system is presented in Figure 6. In this system, the MIR signal is incident onto the hybrid metasurface filter. The transmitted signal is subsequently introduced into a gas cell, which is connected to a detector. This detector measures the intensity of the signal both before and after the introduction of $CO_2$, enabling the detection of $CO_2$. NDIR systems detect changes in IR optical absorption, and the utilization of different absorption bands in the IR enables selectivity to a particular target gas. $CO_2$ strongly absorbs IR radiation at wavelengths around 4.3 µm and has another absorption band around 2.6 µm, which was used for sensing in this study. It is worth mentioning that the absorption level at 2.6 µm is much lower than that of 4.3 µm [63-65].

Beer-Lambert's Law [66] can be used to calculate the intensity of the detected optical signal $I(c)$ for an NDIR system and is given by $I(c) = I_o exp^{(-k_g c l)}$ where $I_o$ is the optical intensity of the source, $c$ is the concentration of the gas, $k_g$ is the absorption index of $CO_2$ at the detection wavelength, and $l$ is the optical path length between the source and detector. The intensity of IR radiation decreases exponentially with an increase in path length. A shorter path length increases the signal output from the IR detector, however, the sensitivity to lower gas concentrations is decreased.

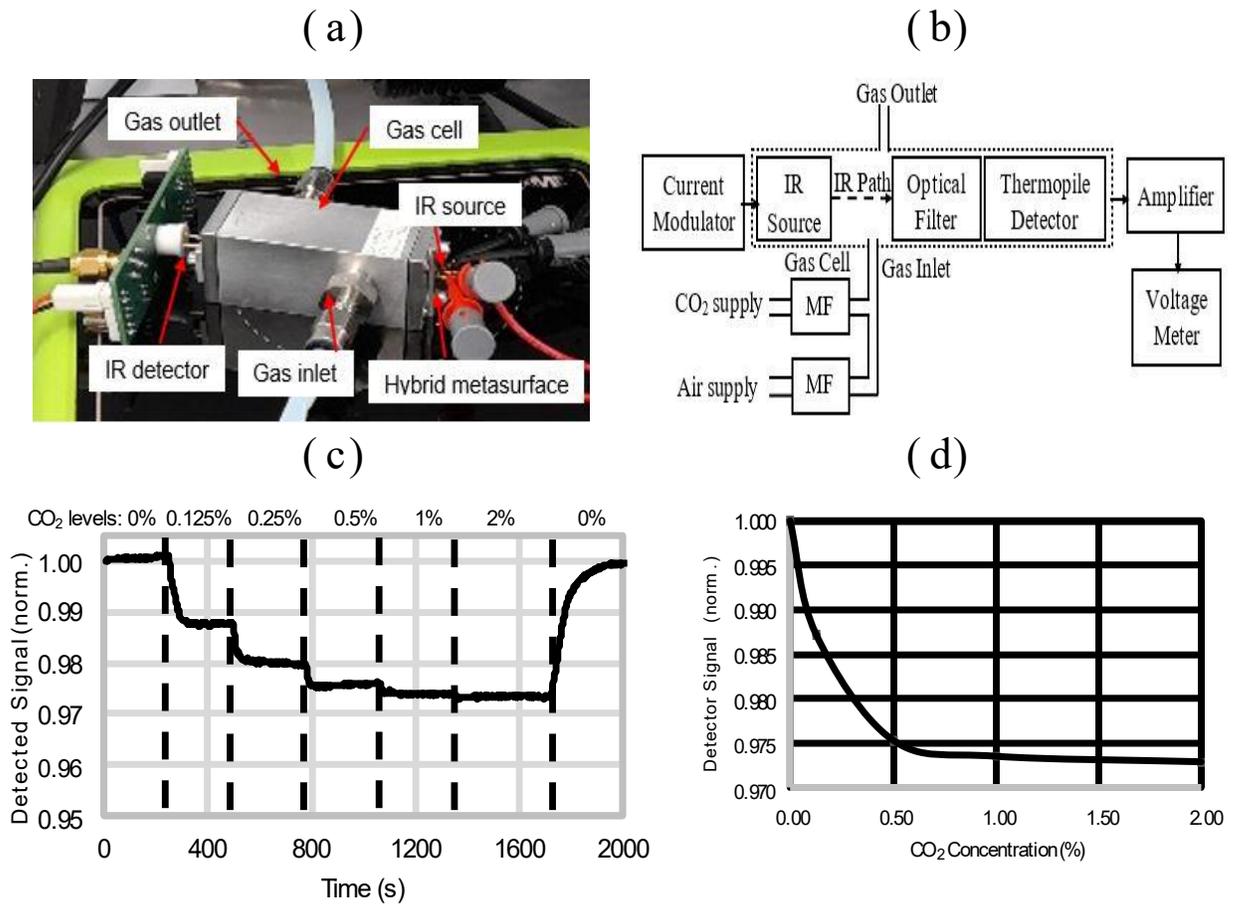

Figure 7. (a) NDIR gas sensor test setup (b) Functional diagram of the NDIR sensor and gas test systems (c) Sensor outputs (relative to the baseline of synthetic air) with $CO_2$ in the range of 0.25% to 2 % (d) Detected signal vs. $CO_2$ concentration from 0.25 % – 2 %.

The NDIR system comprises a wideband IR micro-hotplate-based thermal source and a single channel silicon thermopile detector (HMS-J21, Heimann Sensor, Germany) with an active area of 1.2 mm × 2 mm and a noise level of 37 nV/√Hz and specific detectivity of $1.4 \times 10^8$ cm.√Hz/W. The IR source was previously described [67] and has been developed to have excellent wideband emissivity using a 'black' nanotube layer and can operate at temperatures above 600°C. The Al-Ge metasurface was mounted on the optical window of the IR detector. The components were mounted in an optical gas cell constructed from an Al block with a path length of 55 mm and are shown in Figure 7(a). Gas was injected into the system from a custom-built automated gas control system comprising a set of mass flow controllers (MFCs), enabling the level of $CO_2$ to be varied by balancing with synthetic zero-grade air. $CO_2$ was obtained from a calibration cylinder with a 5 % concentration in air. The station is controlled via a LabVIEW© interface. A block diagram of the system is shown in Figure 7(b).

The received signal from the thermopile detector is in the micro-volt range and was amplified by a 'zero drift' programmable gain differential amplifier (TI LMP8358) with a voltage gain setting of 100x. The signal was low pass filtered and digitized by a Keithley 2401 source meter. The IR source was modulated in constant current mode at a frequency of 1 Hz with a peak temperature of 570 °C at 206 mW using a second Keithley source meter. Changes in the amplitude of the detected signal correspond to the concentration of $CO_2$ to which the sensor is exposed.

The sensor was exposed to $CO_2$ concentrations from 0.25 % to 2 %, with a constant flow rate of 0.2 SLPM. The results in Figure 7(c) show the signal's amplitude normalized to a baseline of one. As the $CO_2$ absorbs the IR radiation, the detected signal decreases with an increase in the concentration of $CO_2$. The normalized response is also plotted as a function of $CO_2$ concentration in Figure 7(d). The limit of detection (LOD) of the sensor system is ~0.04 % $CO_2$, which corresponds to several hundred ppm and is within the sensitivity requirements for a large variety of commercial gas sensing applications. The detection limit can be optimized in the future through signal processing alterations, surface roughness reduction, and tuning the filter to transmit stronger $CO_2$ absorption bands at 4.2 µm, 4.28 µm or 4.3 µm.

**Conclusion**

Hybrid dielectric-metal metasurfaces combine the advantages of both plasmonic metals and dielectric metasurfaces, offering significant potential as high transmission MIR optical filters. Here we demonstrated a bandpass filter targeted to the 2.6 um $CO_2$ absorption mode with a high transmission efficiency of 80% and a narrow FWHM of 0.4 µm. The metasurface was successfully integrated into a prototype gas sensing system, achieving a LOD of ~0.04%. We presented numerical simulation results, a straightforward nanofabrication process based on a single lithography step using SML EBL resist. The fabricated metasurfaces were optically characterized using a microscopic FTIR showing a good agreement with the simulation results. Our results highlight the potential for the of metal-dielectric metasurfaces as customizable optical filters in MIR sensing and spectroscopy applications.

**Notes**

The authors declare no competing financial interest.

**ACKNOWLEDGMENTS**

This work was funded by the ISDB Cambridge Trust. This work was also supported by the Henry Royce Institute for advanced materials through the Equipment Access Scheme enabling access to the Electron-Beam Lithography (Royce EBL Suite) at Cambridge; Cambridge Royce facilities grant EP/P024947/1 and Sir Henry Royce Institute - recurrent grant EP/R00661X/1". Additionally, the work was supported by Rank Prize Institute.